\documentclass[preprint,12pt,authoryear]{elsarticle}

\usepackage{graphicx} 
\usepackage{amsmath}
\usepackage[letterpaper,top=2cm,bottom=2cm,left=3cm,right=3cm,marginparwidth=1.75cm]{geometry}

\usepackage[colorlinks=true, allcolors=blue]{hyperref}

\usepackage{amssymb}

\journal{Journal of Computational Physics}

\begin{document}

\begin{frontmatter}



\title{High-order stochastic integration schemes for the Rosenbluth-Trubnikov collision operator in particle simulations}


\author[inst1]{Zhixin Lu}
\author[inst1]{Guo Meng}
\author[inst2,inst1,inst3]{Tomasz Tyranowski}
\author[inst1]{Alex Chankin}

\affiliation[inst1]{organization={Max Planck Institute for Plasma Physics},
            addressline={Boltzmannstr. 2}, 
            city={Garching},
            postcode={85748}, 
            country={Germany}}
\affiliation[inst2]{organization={University of Twente, Department of Applied Mathematics},
            addressline={PO~Box~217}, 
            city={Enschede},
            postcode={7500AE}, 
            country={Netherlands}}
\affiliation[inst3]{organization={Technical University of Munich, Department of Mathematics},
            addressline={Boltzmannstr.~3}, 
            city={Garching},
            postcode={85748}, 
            country={Germany}}

\begin{abstract}
In this study, we consider a numerical implementation of the nonlinear Rosenbluth-Trubnikov collision operator for particle simulations in plasma physics in the framework of the finite element method (FEM). The relevant particle evolution equations are formulated as stochastic differential equations, both in the Stratonovich and It\^o forms, and are then solved with advanced high-order stochastic numerical schemes. Due to its formulation as a stochastic differential equation, both the drift and diffusion components of the collision operator are treated on an equal footing. Our investigation focuses on assessing the accuracy of these schemes. Previous studies on this subject have used the Euler-Maruyama scheme, which, although popular, is of low order, and requires small time steps to achieve satisfactory accuracy. In this work, we compare the performance of the Euler-Maruyama method to other high-order stochastic methods known in the stochastic differential equations literature. Our study reveals advantageous features of these high-order schemes, such as better accuracy and improved conservation properties of the numerical solution. The main test case used in the numerical experiments is the thermalization of isotropic and anisotropic particle distributions.
\end{abstract}





\end{frontmatter}



\section{Introduction}
The collision operator in gyrokinetic/kinetic simulations takes into account important physics such as neoclassical transport \cite{hinton1976theory,wang2006nonlocal}, the damping of the zonal flow \cite{lin1999effects}, the evolution of the Alfv\'en modes \cite{chen1997collisional}, and the collision transport in the edge plasma \cite{chankin2012development}. Up to now, most of the particle codes treat the collision operator using the Langevin equation, and the convergence order is limited to the widely used Euler-Maruyama scheme \cite{xu1991numerical,lin1999effects,donnel2020moment} or the splitting scheme \cite{slaby2017combining}, both with a global first-order weak convergence. Even though a noise reduction scheme has been proposed \cite{sonnendrucker2015split}, the time step size has to be very small when using the Euler-Maruyama scheme or other low-order schemes  to achieve the desired accuracy. 

For particle simulations, the basic equation describing the collision process is the Langevin equation, that is, a type of stochastic differential equation (SDE) driven by a Wiener process. The theoretical and numerical aspects of SDEs have been extensively studied in the last several decades, and a vast number of numerical integration techniques have been proposed \cite{ito1944109,stratonovich1966new,kloeden1992stochastic}, for instance, various higher-order stochastic Runge-Kutta schemes for Stratonovich SDEs, both strongly (\cite{burrage1996high,burrage2000order}) and weakly (\cite{rossler2007second}) convergent. Previous work related to high-order stochastic schemes for collision operators is limited to the Milstein-like scheme \cite{dimits2013higher}. More recent work includes the development of stochastic variational principles for collisional kinetic equations \cite{tyranowski2021stochastic}, and variational integrators for stochastic dissipative Hamiltonian systems \cite{kraus2021variational}.

In this work, we focus on the application of high-order stochastic integration schemes to the Langevin equation associated with the Fokker-Planck equation with a collision operator defined by the Rosenbluth-Trubnikov potentials \cite{trubnikov1965particle,rosenbluth1957fokker}. The scope and goal of this paper are as follows. 
\begin{enumerate}
    \item We aim for the identification of the different features related to the particle simulations such as the particle noise and the representation of the distribution using particles, in addition to the traditional analyses of the SDE studies \cite{kloeden1992stochastic,burrage1996high}. 
    \item One goal is to evaluate the role of high-order stochastic schemes in the improvement of the overall accuracy of the collision operator in particle simulations, compared with the Euler-Maruyama scheme that is widely used in the present gyrokinetic particle codes \cite{slaby2017combining,lanti2020orb5,lin1999effects}.
    \item Our main intention is for the full $f$ particle simulations, motivated by recent development of full $f$ gyrokinetic code \cite{hager2016fully,lu2023full}. A specific further application will be that in TRIMEG code \cite{lu2019development,lu2021development,lu2023full}. In addition, high-order stochastic integration schemes for the collision operator are also expected to bring benefits to the gyrokinetic $\delta f$ particle codes \cite{lanti2020orb5,slaby2017combining}.
\end{enumerate} 
The rest of the paper is organized as follows. In Sec. \ref{sec:model}, the physics model for the collision operator is given. In Sec. \ref{sec:sde_integrator}, the time integration schemes are listed. In Sec. \ref{sec:results}, the application of various schemes to the collision operator is presented, accompanied by the numerical results. The conclusion and outlook are given in Sec. \ref{sec:summary}. 

\section{Models and Equations}
\label{sec:model}

\subsection{General equations of collision operator}

We adopt the Rosenbluth-Trubnikov collision operator \cite{trubnikov1965particle,rosenbluth1957fokker}. The kinematics of the collision between a particle of type $a$ with velocity $\mathbf{v}$ and a particle of type $b$ with velocity $\mathbf{v}'$ are discussed. The Fokker-Planck equation with collisions, accounting for particles of type $a$ with distribution function $f_a$, can be written as follows,
\begin{eqnarray}
    \label{eq: Fokker-Planck equation with a collision operator}
    \left(\frac{\partial f_a}{\partial t}\right)_{coll} = C(f_a,f_b) \;\;,
\end{eqnarray}
where $C(f_a,f_b)$ is the collision operator. In this work, we use the Rosenbluth-Trubnikov potentials when calculating the collisions, namely, 
\begin{eqnarray}
    C(f_a,f_b) = -\Gamma^{a\backslash b} \frac{\partial}{\partial\mathbf{v}}\cdot \left [f_a \frac{\partial}{\partial\mathbf{v}} \hat h(\mathbf{v}) \right] 
    +\frac{\Gamma^{a\backslash b}}{2}  \frac{\partial}{\partial\mathbf{v}}\cdot  \frac{\partial}{\partial\mathbf{v}}\cdot \left [f_a \frac{\partial}{\partial\mathbf{v}} \frac{\partial}{\partial\mathbf{v}} \hat g(\mathbf{v}) \right]  \;\;,
\end{eqnarray}
where $\hat g$ and $\hat h$ are the Rosenbluth-Trubnikov potentials \cite{rosenbluth1957fokker}, $\Gamma^{a\backslash b}=4\pi\lambda_{ab}e_a^2e_b^2/m_a^2$ in CGS units or $\Gamma^{a\backslash b}=\lambda_{ab}e_a^2e_b^2/(4\pi\varepsilon_0^2 m_a^2)$ in SI units, $\varepsilon_0$ is the vacuum permittivity, $\lambda_{ab}$ is the Coulomb logarithm, $e_a$ and $e_b$ are the electric charges, and $m_a$ and $m_b$ are the masses of the particles of species $a$ and $b$, respectively. 
The Rosenbluth-Trubnikov potentials are as follows,
\begin{eqnarray}
    \hat h &=&\frac{m_a+m_b}{m_b} h \;\;, \\
    \hat g &=& g \;\;, \\
\label{eq:h_integral}
    h(\mathbf{v}) &=& \int d\mathbf{v}' \frac{f_b(\mathbf{v}')}{|\mathbf{v}-\mathbf{v}'|} \;\;, \\
\label{eq:g_integral}
    g(\mathbf{v}) &=& \int d\mathbf{v}' f_b (\mathbf{v}')|\mathbf{v}-\mathbf{v}'| \;\;.
\end{eqnarray}
The elliptic equations are satisfied as follows,
\begin{eqnarray}
    \frac{\partial}{\partial\mathbf{v}}\cdot \frac{\partial}{\partial\mathbf{v}} h(\mathbf{v}) &=& -4\pi f_b(\mathbf{v}) \;\;, \\
     \frac{\partial}{\partial\mathbf{v}}\cdot \frac{\partial}{\partial\mathbf{v}} g(\mathbf{v}) &=& 2 h(\mathbf{v}) \;\;. 
\end{eqnarray}
In the following, the subscript `$a$' is omitted when no ambiguity is introduced.  

\subsection{Discretization of the distribution function}
The particle distribution $f$ is represented by the so-called markers which are the numerical particles in simulations. Take the phase space coordinates $\bf z$, and the marker distribution is given by $f_{mark}({\bf z})$ using the Klimontovich representation,
\begin{eqnarray}
    f_{mark}^{Klim}({\bf z}) = \sum_{i=1}^{N_p} \frac{\delta({\bf z}-{\bf z}_i)}{J_z} \;\;,
\end{eqnarray}
where $\delta(x)$ is a Dirac delta function, $J_z$ is the Jacobian of the coordinates $\bf z$, ${\bf z}=({\bf R, v} )$, $\bf R$ is the configuration space coordinate and $\bf v$ is the velocity.
The physical distribution function is represented by the markers whose coordinates are $\bf z_p$ and given by
\begin{eqnarray}
\label{eq:f_discretization}
    f^{Klim}({\bf z}) = C_{p2g} \sum_{i=1}^{N_p} p_i \frac{\delta({\bf z}-{\bf z}_i)}{J_z} \;\;,
\end{eqnarray}
where $p_i=f({\bf z}_i)/[f_{mark}({\bf z}_i)C_{p2g}]$, $C_{p2g}=N_{phy}/N_p$, $N_p$ is the number of markers, $N_{phy}$ is the number of physical particles, $p_i$ is the full $f$ weight of marker `$i$'. The variable $C_{p2g}$ is defined so that $p_i=1$ if the marker distribution is chosen to be proportional to the physical particle distribution. 

Since the noise exists due to finite marker number, Eq. \ref{eq:f_discretization} is not exactly the desired analytically given distribution function $f_A({\bf z})$. The relative error can be defined to describe the deviation of the numerical distribution $f$ and the analytical distribution $f_A$. One possible definition is as follows.
A complete orthonormal set of functions $B_{\bf I}({\bf z})$ is chosen to represent any functions in the phase space $\bf z$,
\begin{eqnarray}
\label{eq:orthogonal}
    \int d{\bf z} B_{\bf I}({\bf z})  B_{{\bf I}'}({\bf z}) &=& \delta_{\bf I,I'} \;\;,
\end{eqnarray}
where $\delta_{\bf I,I'}=0$ if ${\bf I}\ne {\bf I}'$ and $\delta_{\bf I,I'}=1$ if ${\bf I}= {\bf I}'$.
Then the discretized distribution $f({\bf z})$ and the analytical distribution $f_A({\bf z})$ are expressed as follows,
\begin{eqnarray}
\label{eq:fz_decomp}
    f({\bf z}) &=& \sum_{\bf I}f_{\bf I} B_{\bf I}({\bf z})  \;\;, \\
    f_A({\bf z}) &=& \sum_{\bf I}f_{A,{\bf I}} B_{\bf I}({\bf z})  \;\;,
\end{eqnarray}
The coefficients $f_{\bf I}$ can be obtained from the weak form of Eq. \ref{eq:fz_decomp},
\begin{eqnarray}
    \sum_{\bf I} \int d{\bf z} B_{\bf I} B_{\bf I'} f_{\bf I} =\int d{\bf z} B_{\bf I'}({\bf z}) f({\bf z}) \;\;, 
\end{eqnarray}
whose solution is readily obtained considering Eq. \ref{eq:orthogonal},
\begin{eqnarray}
     f_{\bf I} =\int d{\bf z} B_{\bf I}({\bf z}) f({\bf z}) \;\;.
\end{eqnarray}
Similarly, 
\begin{eqnarray}
     f_{A,{\bf I}} =\int d{\bf z} B_{\bf I}({\bf z}) f_A({\bf z}) \;\;.
\end{eqnarray}
The relative error is
\begin{eqnarray}
    \epsilon_f \equiv \sqrt{ \frac{\int d{\bf z} [ f({\bf z})-f_A({\bf z}) ]^2}{\int d{\bf z}  f_A^2({\bf z}) } }
    = \sqrt{ \frac{\sum_{\bf I} (f_{\bf I}-f_{A,{\bf I}})^2}{\sum_{\bf I}f_{A,{\bf I}}^2}}\;\;.
\end{eqnarray}
A more convenient way as we adopted in the numerical studies is to calculate the relative error of the kinetic energy, 
\begin{eqnarray}
\label{eq:relative_errorE}
    \epsilon_E =  \frac{C_{p2g} \sum_{i=1}^{N_p} p_i v_i^2}{\int d {\bf z} f_A({\bf z})v^2 }-1  \;\;.
\end{eqnarray}
For Maxwellian distribution $f_A=[n/(\pi 2T/m)^{3/2}]\exp\{-mv^2/(2T)\}$, we can readily get  $\int d {\bf z} f_A({\bf z})v^2=3nT/m=(3/2)nv_t^2$, where $v_t=\sqrt{2T/m}$ is the thermal velocity.

\subsection{Drag and diffusion coefficients in the Langevin equation}
In the following work, we adopt the drift kinetic model for which magnetically confined plasmas are considered. In velocity space, $(v_\|,v_\perp,\alpha)$ is adopted where $v_\|, v_\perp$ are the components of velocity in the directions parallel and perpendicular to the magnetic field, and $\alpha$ is the gyro angle. The Jacobian of $(v_\|,v_\perp,\alpha)$ is $J_v=v_\perp$. It is assumed that the distribution and the Rosenbluth-Trubnikov potentials are independent of $\alpha$.

The Langevin equation is solved  in $(v_\|,y)$ coordinates, where $y=v_\perp^2$, instead of $(v_\|,v_\perp)$ so that most of the singular terms $1/v_\perp$ can be eliminated. Specifically, the collision operator is applied by modifying the markers' $(v_\|,y)$. Note that the diffusion coefficients are still calculated in $(v_\|,v_\perp)$ coordinates since the Rosenbluth-Trubnikov potentials are solved in $(v_\|,v_\perp)$ coordinates, for the sake of simplicity. Specifically, in solving the Rosenbluth-Trubnikov potentials using particles of arbitrary distribution, the Neumann boundary condition can be applied at $v_\perp=0$. 

The collision operator is written as
\begin{eqnarray}
\label{eq:langevin_dfdt}
    \partial_t f|_c 
    &=& \Gamma \left[
    \partial_x (D_x f) +\partial_y (D_y f)
    +\partial^2_{xx}(D_{xx}f) +\partial^2_{xy}(D_{xy}f)+\partial^2_{yy}(D_{yy}f) \right] \\
\label{eq:langevin_drag}
    D_x &=& -\frac{m_a}{m_b} \partial_\| h \;\;, 
    D_y   =  -2\frac{m_a}{m_b} v_\perp\partial_\perp  h -\partial_{\perp\perp}^2g-\frac{1}{v_\perp} \partial_\perp g \;\;, \;\; \\
\label{eq:langevin_diffusion}
    D_{xx} &=& \frac{1}{2} \partial^2_{\|\|} g \;\;, \;\;
    D_{xy} =D_{yx} = v_\perp \partial^2_{\|\perp} g \;\;, \;\;
    D_{yy}  = 2v_\perp^2 \partial^2_{\perp\perp} g \;\;, \;\;
\end{eqnarray}
where the subscripts `$\|$' and `$\perp$' indicate the directions parallel and perpendicular to the magnetic field respectively, $\partial_\perp\equiv\partial/\partial v_\perp$, $\partial_\|\equiv\partial/\partial v_\|$, $h$, $g$ and their derivatives are calculated in $(v_\|,v_\perp)$ coordinates but the diffusion operator is written in $(x,y)$ coordinates, where $x=v_\|$, $y=v_\perp^2$. It can be shown that Eqs. \ref{eq:langevin_dfdt}--\ref{eq:langevin_diffusion} are equivalent to those in the continuum code KIPP \cite{chankin2012development} implemented completely in $(v_\|,v_\perp)$.

Since Eq.~\ref{eq: Fokker-Planck equation with a collision operator} is a Fokker-Planck equation, the particle density function $f$ can be interpreted as the probability density for a $d$-dimensional stochastic process ${\bf v}(t)$ which satisfies the It\^o SDE \cite{kloeden1992stochastic} 
\begin{eqnarray}
    \label{eq: Ito SDE}
    d{{\bf v}} &=& {\bf F}_{Ito}  dt + \overline{\overline{G}} \cdot d{\bf W} \;\;,
\end{eqnarray}
driven by the standard $m$-dimensional Wiener process ${\bf W}(t)$, where ${\bf F}_{Ito}$ is the drag (or drift) function, and $\overline{\overline{G}}$ is the $d \times m$ diffusion matrix. The dimensions of the ${\bf v}(t)$ and ${\bf W}(t)$ processes are in our case $d=2$ and $m=2$, respectively. We further have

\begin{eqnarray}
\label{eq:drag4Ito}
    {\bf F}_{{Ito} } &=& - \begin{bmatrix} D_x \\ D_y \end{bmatrix} \;\;, \\
     \overline{\overline{G}}\cdot  \overline{\overline{G}}^T  &=&  2\overline{\overline{D}} 
    = 2\begin{bmatrix}
      D_{xx} & D_{xy} \\ 
      D_{yx} & D_{yy} \\
    \end{bmatrix} \;\;,
\end{eqnarray}
and with the choice $G_{12}=0$, we can calculate the diffusion matrix as 
\begin{eqnarray}
\label{eq:matG}
     \overline{\overline{G}}
    = \sqrt{2}
    \begin{bmatrix}
      \sqrt{D_{xx}} & 0 \\ 
      \frac{D_{yx}}{\sqrt{D_{xx}}} & \sqrt{D_{yy}-\frac{D_{x y}^2}{D_{xx}}} \\
    \end{bmatrix} \;\;.
\end{eqnarray}
Independent realizations (or sample paths) of the stochastic process ${\bf v}(t)$ can be interpreted as the trajectories of the markers in the discretization Eq. \eqref{eq:f_discretization}.

The It\^o SDE Eq. \eqref{eq: Ito SDE} can also be recast in the Stratonovich form

\begin{eqnarray} 
    \label{eq: Stratonovich SDE}
    d{{\bf v}} &=& {\bf F}_{Str}  dt + \overline{\overline{G}} \circ d{\bf W} \;\;,
\end{eqnarray}
where the drag function ${\bf F}_{Str}$ is related to ${\bf F}_{{Ito} }$ through the Stratonovich correction formula

\begin{eqnarray}
\label{eq:Ito2Stra}
    {F}^i_{Str}  = {F}^i_{Ito}  -\frac{1}{2}\sum_{j=1}^m\sum_{k=1}^d G^{kj} \frac{\partial G^{ij}}{\partial v^k} \;\;,
\end{eqnarray}
for $i=1,\ldots,d$. Substituting Eqs. \eqref{eq:drag4Ito} and \eqref{eq:matG} into Eq. \eqref{eq:Ito2Stra}, we get
\begin{eqnarray}
\label{eq:drag4stra}
    {\bf F}_{Str}  &=& \begin{bmatrix}  
        -D_x-\frac{1}{2}\left(
            \partial_x D_{xx} +
            \frac{D_{x y}}{D_{xx}}\partial_y D_{xx}
        \right)\\
        -D_y-\frac{1}{2}  \left(
            -\frac{D_{x y}}{D_{xx}}\partial_x D_{xx} +
            2\partial_x D_{x y} 
            +\partial_y D_{yy}
        \right)
    \end{bmatrix},
\end{eqnarray}
where 
\begin{eqnarray}
\label{eq:dDxxdx}
    \partial_xD_{xx}&=&\frac{1}{2}\partial^3_{\|\|\|}g \;\;, \\
    \partial_yD_{xx}&=&\frac{1}{4v_\perp}\partial^3_{\|\|\perp}g \;\;, \\
    \partial_xD_{xy}&=&v_\perp\partial^3_{\|\|\perp}g \;\;,  \\
    \partial_yD_{xy}&=&\frac{1}{2}\partial^3_{\|\perp\perp}g +\frac{1}{2v_\perp}\partial^2_{\|\perp}g\;\;, \\
    \partial_xD_{yy}&=&2v_\perp^2\partial^3_{\|\perp\perp}g \;\;,  \\
\label{eq:dDyydy}
    \partial_yD_{yy}&=&v_\perp\partial^3_{\perp\perp\perp}g+2\partial^2_{\perp\perp}g \;\;. 
\end{eqnarray}

Depending on the choice of It\^o or Stratonovich forms, the equations to be solved are listed in Table \ref{tab:ito_stra}, where the numerical integrators adopted in the numerical studies in Section \ref{sec:sde_integrator} are also listed.

\begin{table}[h!]
\centering
\begin{tabular}{|c |c |c| c |} 
 \hline
  Representation &  Drag & Diffusion & Integrator  \\ 
  \hline
  It\^o & Eqs. \ref{eq:drag4Ito}, \ref{eq:langevin_drag}  
  &  Eqs. \ref{eq:matG}, \ref{eq:langevin_diffusion} 
  & Euler-Maruyama, Splitting\\ 
  \hline
  Stratonovich & Eqs. \ref{eq:drag4stra}, \ref{eq:langevin_drag}  
  & Eqs. \ref{eq:matG}, \ref{eq:langevin_diffusion}   
  & Heun, PL, CL, E1, G5\\ 
  \hline
\end{tabular}
\caption{Equations for the It\^o and Stratonovich representations.}
\label{tab:ito_stra}
\end{table}

We say that the noise driving an SDE is \emph{commutative}, when the diffusion matrix satisfies the condition \cite{kloeden1992stochastic}
\begin{eqnarray}
    \sum_{k=1}^d G_{kj}\frac{\partial}{\partial v_k} G_{ri} = 
    \sum_{k=1}^d G_{ki}\frac{\partial}{\partial v_k} G_{rj} \;\;,
\end{eqnarray}
for all $i,j=1,\ldots,m$, and $r=1,\ldots,d$. It can be easily verified that the diffusion matrix \eqref{eq:matG} does not satisfy this condition, therefore the noise in Eqs.~\ref{eq: Ito SDE} and \eqref{eq: Stratonovich SDE} is \emph{non-commutative}.

\subsection{The mixed-MC-FEM solution for the Rosenbluth-Trubnikov potentials}
\label{subsec:MC_FEM_equation}
In the interior regime of $(v_\|,v_\perp)$ space, the Rosenbluth-Trubnikov potentials are solved from the elliptic equations.
Let us consider a piecewise smooth regularization $f$ of the particle distribution $f^{Klim}$ in Eq. \ref{eq:f_discretization}, represented by the cubic spline using the finite element method (FEM),
\begin{equation}
    f(\mathbf{R},v_\|,v_\perp,\alpha)
    =\sum_{i_\|,i_\perp}f_{i_\|,i_\perp}(\mathbf{R})
    N_{i_\|}(v_\|) N_{i_\perp} (v_\perp) \;\;,
\end{equation}
where $N_{i_\|}$, $N_{i_\perp}$ are the basis functions, and symmetry in the $\alpha$ direction is assumed.
We only treat the distribution in velocity space and omit the configuration coordinates in the following studies for the sake of simplicity. The dependency on the configuration space can be taken into account readily when it is needed in further studies. 
The weak form for the coarse distribution is obtained as follows,
\begin{eqnarray}
    \int dv_\|dv_\perp d\alpha J_v f    N_{i_\|}(v_\|) N_{i_\perp} (v_\perp) 
    &=& \int dv_\|dv_\perp d\alpha J_v f^{Klim}   N_{i_\|}(v_\|) N_{i_\perp} (v_\perp)\;\;,
\end{eqnarray}
which yields, 
\begin{eqnarray}
\label{eq:Mf=rhs}
    \overline{\overline{M}}_{p2g,i_\|,i_\perp,j_\|,j_\perp} f_{j_\|,j_\perp}
    &=& \sum_{i=1}^{N_p} p_i  N_{i_\|}(v_{\|,i}) N_{i_\perp} (v_{\perp,i})  \;\;, \\
    \overline{\overline{M}}_{p2g,i_\|,i_\perp,j_\|,j_\perp} 
    &=&
    2\pi \int dv_\|dv_\perp J_v   N_{i_\|}(v_\|) N_{i_\perp} (v_\perp)   N_{j_\|}(v_\|) N_{j_\perp} (v_\perp) \;\;.
\end{eqnarray}
In our approach, the coefficients $f_{i_\|,i_\perp}$ are not computed when calculating the Rosenbluth-Trubnilov potentials. Thus the matrix inversion in Eq. \ref{eq:Mf=rhs} is not needed unless $f_{j_\|,j_\perp}$ is needed for other purposes such as diagnostics. In this sense, our approach differs from the mixed-particle-grid schemes whose collision is based on $f$ representation using grids \cite{hager2016fully}. 

The elliptic equations in $(v_\|,v_\perp,\alpha)$ coordinates are as follows,
\begin{eqnarray}
    \left[ 
    \frac{\partial^2 }{\partial v_\|^2}  +\frac{1}{v_\perp}  \frac{\partial }{\partial v_\perp} v_\perp  \frac{\partial }{\partial v_\perp} 
    \right] h(v_\|,v_\perp)
    &=& -4\pi f(v_\|,v_\perp) \;\;,\\
    \left[ 
    \frac{\partial^2 }{\partial v_\|^2}  +\frac{1}{v_\perp}  \frac{\partial }{\partial v_\perp} v_\perp  \frac{\partial }{\partial v_\perp} 
    \right] g(v_\|,v_\perp)
    &=& 2 h(v_\|,v_\perp) \;\;.
\end{eqnarray}
With the scalar product $\int dv_\| dv_\perp d\alpha  J_v N_{i_\|}(v_\|) N_{i_\perp} (v_\perp) \ldots$, we obtain the weak form
\begin{eqnarray}
\label{eq:elliptic_h}
    \overline{\overline{M}}_{h,i_\|,i_\perp,j_\|,j_\perp} h_{j_\|,j_\perp}
    &=& -4\pi\sum_{i=1}^{N_p} p_i  N_{i_\|}(v_{\|,i}) N_{i_\perp} (v_{\perp,i})  \;\;, \\
    \overline{\overline{M}}_{h,i_\|,i_\perp,j_\|,j_\perp} 
    &=&
    -2\pi \int dv_\|dv_\perp  v_\perp
    \left[
     \frac{\partial N_{i_\|}}{\partial v_\|}  N_{i_\perp}  \frac{\partial N_{j_\|}}{\partial v_\|}  N_{j_\perp} 
    +N_{i_\|} \frac{\partial N_{i_\perp}}{\partial v_\|}   N_{j_\|} \frac{\partial N_{j_\perp}}{\partial v_\perp}
    \right]
    \;\;.
\end{eqnarray}
The weak form equation for $g$ is as follows,
\begin{eqnarray}
\label{eq:elliptic_g}
    \overline{\overline{M}}_{g,i_\|,i_\perp,j_\|,j_\perp}^L g_{j_\|,j_\perp}
    &=&  \overline{\overline{M}}_{g,i_\|,i_\perp,j_\|,j_\perp}^R h_{j_\|,j_\perp}
    \;\;, \\
    \overline{\overline{M}}_{g,i_\|,i_\perp,j_\|,j_\perp} ^L
    &=& 
    \overline{\overline{M}}_{h,i_\|,i_\perp,j_\|,j_\perp} \;\;, \\
    \overline{\overline{M}}_{g,i_\|,i_\perp,j_\|,j_\perp} ^R
    &=&
      2\int dv_\|dv_\perp  v_\perp
     N_{i_\|} N_{i_\perp} N_{j_\|} N_{j_\perp}
    \;\;.
\end{eqnarray}

The Monte-Carlo (MC) integration is adopted for the calculation of the values of $h$ and $g$ along the boundaries, using the markers as the samplers. There are fours edges in $(v_\|,v_\perp)$ space as the boundary, namely,
\begin{itemize}
    \item $v_\perp=v_{\perp,min}, v_\|\in [v_{\|,min},v_{\|,max}]$ ,
    \item $v_\perp=v_{\perp,max}, v_\|\in [v_{\|,min},v_{\|,max}]$ ,
    \item $v_\|=v_{\|,min}, v_\perp\in [v_{\perp,min},v_{\perp,max}]$ ,
    \item $v_\|=v_{\|,max}, v_\perp\in [v_{\perp,min},v_{\perp,max}]$ ,
\end{itemize}
where $v_{\perp,min}=0$, $v_{\|,min}=-v_{\|,max}$, $v_{\perp,max}$ and $v_{\|,max}$ are at least $3$ times of the thermal velocity. It can be readily proved that in $(v_\|,v_\perp)$ coordinates, the boundary condition at $v_\perp=v_{\perp,min}$ for the Rosenbluth-Trubnikov potentials is 
\begin{eqnarray}
\label{eq:neumann_h}
    \partial_{v_\perp} h (v_\perp=v_{\perp,min}) = 0\;\;, \\
\label{eq:neumann_g}
    \partial_{v_\perp} g (v_\perp=v_{\perp,min}) = 0\;\;.
\end{eqnarray}
As a result, the Neumann boundary condition is adopted at $v_\perp=v_{\perp,min}$.
The Dirichlet boundary condition is adopted in the other three edges where the values of $g$ and $h$ are calculated from the integral form in Eqs. \ref{eq:h_integral} and \ref{eq:g_integral}.
Along the boundary where the Dirichlet boundary condition is applied,  the direct calculation of $h$ and $g$ is adopted,
\begin{eqnarray}
\label{eq:h_pic}
    h(\mathbf{v}) 
    &=& 4 \int\int dv_\|' dy' J_y \frac{1}{u} f(v_\|',y') K(k) \;\;, \nonumber\\
    &=& \frac{2}{\pi}\int\int dv_\|' dy' \int d\alpha' J_y \frac{1}{u} f(v_\|',y') K(k) \;\;, \nonumber \\
    &=& \frac{2}{\pi} \sum_{i=1}^{N_p} 
    p_i\left[
    \frac{1}{u}  K(k) 
    \right]_i \;\;,  \\
\label{eq:g_pic}
    g(\mathbf{v}) &=&  \frac{2}{\pi} \sum_{i=1}^{N_p} 
    p_i\left[
    u  E(k) 
    \right]_i \;\;, \\
     u&=&\sqrt{(v_\|-v_\|')^2+(v_\perp+v_\perp')^2} \;\;, \\
    k&=&2\frac{\sqrt{v_\perp v_\perp'}}{u} \;\;,
\end{eqnarray}
where $[\ldots]_i$ indicates that the value is evaluated at the location of marker `$i$', $J_y=1/2$, $K(k)$, and $E(k)$ are the complete elliptic integrals of the first and second kinds, respectively. In deriving Eqs. \ref{eq:h_pic} and \ref{eq:g_pic}, the representation of $f$ in Eq. \ref{eq:f_discretization} using the markers is adopted. 

\subsection{Test particle collisions due to Maxwellian background}
While the general form of the Rosenbluth-Trubnikov potentials can be obtained as formulated in Section \ref{subsec:MC_FEM_equation}, the collision operator due to the Maxwellian background particles is widely used in studies of energetic particle \cite{meng24NF} and electron neoclassical transport \cite{lin1995gyrokinetic}. Since the dimensions of $g$ and $h$ are $nv$ and $n/v$ respectively, in the following, $g$ and $h$ are normalized to $n_bv_{t,b}$ and $n_b/v_{t,b}$ respectively, where $v_{t,b}=\sqrt{2T_b/m_b}$ is the background particle thermal velocity. 
For Maxwellian background, the Rosenbluth-Trubnikov potentials can be calculated analytically \cite{dimits2013higher},
\begin{eqnarray}
\label{eq:h_theory}
    h(u)&=&\frac{\Phi}{u} \;\;, \;\;\\
\label{eq:g_theory}
    g(u)&=&\frac{1}{2}\left[\Phi(2u+\frac{1}{u})+\Phi' \right] \;\;, \\
    \Phi &=& \text{erf}(u) =\int_0^ud\xi \frac{2}{\sqrt{\pi}} e^{-\xi^2} \;\;,
\end{eqnarray}
where $u\equiv v/v_{t,b}$, $v=|\Vec{v}|$, $\Vec{v}$ is the velocity vector. 

The following identities are used for the calculation of the drag/diffusion coefficients.
\begin{eqnarray}
    \partial_u h &=&\left( -\frac{\Phi}{u} + \frac{2}{\sqrt{\pi}} e^{-u^2} \right)\frac{ 1}{u} \;\;, \\
    \partial^2_{uu} h &=& 2\frac{\Phi}{u^3} - \frac{4e^{-u^2}}{\sqrt{\pi}u^2}
    -\frac{4}{\sqrt{\pi}}e^{-u^2} \;\;,  \\
    \partial_u g&=& \frac{1}{2} \left[
    \Phi\left(2-\frac{1}{u^2}\right) +\frac{2}{\sqrt{\pi}u}e^{-u^2}
    \right]   \;\;,\\
    \partial^2_{uu} g&=& 
    \frac{\Phi}{u^3} -\frac{2}{\sqrt{\pi}u^2} e^{-u^2} 
    \;\;, \\
    \partial^3_{uuu}g &= & \left[
      \left(-\frac{3}{u}\Phi +\frac{6}{\sqrt{\pi}} e^{-u^2} \right)\frac{1}{u^2}
      +\frac{4e^{-u^2}}{\sqrt{\pi}} 
    \right]\frac{1}{u} \;\;.
\end{eqnarray}
Since $g$ and $h$ are functions of $u$, the calculation of Eq. \ref{eq:langevin_diffusion} can be simplified considering
\begin{eqnarray}
    \partial_{\|\|}^2g &=& \frac{v_\|^2}{v^2}\partial_{uu}^2g + \frac{v_\perp^2}{v^3}\partial_u g \;\;, \\
    \partial_{\perp\perp}^2g &=& \frac{v_\perp^2}{v^2}\partial_{uu}^2g + \frac{v_\|^2}{v^3}\partial_u g \;\;, \\
    \partial_{\|\perp}^2g &=& \frac{v_\|v_\perp}{v^2}\partial_{uu}^2g - \frac{v_\|v_\perp}{v^3}\partial_u g \;\;, 
\end{eqnarray}
For the drag coefficients in It\^o's form in Eqs. \ref{eq:dDxxdx}--\ref{eq:dDyydy}, the third derivatives of $g$ are needed,
\begin{eqnarray}
    \partial_{\|\|\|}^3g &=& \frac{v_\|^3}{v^3}\partial_{uuu}^3g 
    + \frac{3v_\|v_\perp^2}{v^4}\partial_{uu}^2 g
    -\frac{3v_\|v_\perp^2}{v^5}\partial_u g\;\;, \\
    \partial_{\perp\perp\perp}^3g &=& \frac{v_\perp^3}{v^3}\partial_{uuu}^3g 
    + \frac{3v_\|^2v_\perp}{v^4}\partial_{uu}^2 g
    -\frac{3v_\|^2v_\perp}{v^5}\partial_u g\;\;, \\
    \partial_{\|\perp\perp}^3g &=& \frac{v_\|v_\perp^2}{v^3}\partial_{uuu}^3g 
    + \frac{v_\|^3-2v_\|v_\perp^2}{v^4}\partial_{uu}^2 g
    +\frac{2v_\|v_\perp^2-v_\|^3}{v^5}\partial_u g\;\;, \\
    \partial_{\|\|\perp}^3g &=& \frac{v_\|^2v_\perp}{v^3}\partial_{uuu}^3g 
    + \frac{v_\perp^3-2v_\|^2v_\perp}{v^4}\partial_{uu}^2 g
    +\frac{2v_\|^2v_\perp-v_\perp^3}{v^5}\partial_u g\;\;,
\end{eqnarray}
where $v_\|$, $v_\perp$ and $v$ are normalized to $v_{t,b}$.

\section{High-order integrators for stochastic differential equations (SDE)}
\label{sec:sde_integrator}
In this section we review a number of stochastic Runge-Kutta methods known in the literature  \cite{kloeden1992stochastic,burrage1996high}. Consider a general $N$-dimensional Stratonovich SDE driven by the standard $M$-dimensional Wiener process ($d{\bf W}=[dW_1,dW_2,\ldots,dW_M]^T$),
\begin{eqnarray}
    \label{eq: General Stratonovich SDE}
    d\begin{bmatrix} v_1 \\ v_2 \\ \ldots \\v_N\end{bmatrix}
    &=& \begin{bmatrix} F_1 \\ F_2 \\ \ldots \\F_N \end{bmatrix} dt
    + \begin{bmatrix} 
       G_{1,1}& G_{1,2} & \ldots & G_{1,M} \\ 
       G_{2,1}& G_{2,2} & \ldots & G_{2,M} \\ 
       \ldots \\
       G_{N,1}& G_{N,2} & \ldots & G_{N,M} \\ 
     \end{bmatrix}
     \circ 
     \begin{bmatrix} dW_1 \\ dW_2 \\ \ldots \\dW_M\end{bmatrix} \;\;.
\end{eqnarray}
A general $s$-stage stochastic Runge-Kutta method for Eq. \eqref{eq: General Stratonovich SDE} is defined by the formula \cite{burrage1996high}
\begin{eqnarray}
    V_i &=&v_n+\Delta t \sum_{j=1}^{s} a_{ij} F(V_j) 
    + \sum_{l=1}^p \sum_{j=1}^{s} 
     G(V_j)b_{ij,l}\theta_l, \qquad \text{for $i=1,\ldots,s$} \;\;,\\
    v_{n+1} &=& v_n +\Delta t \sum_{j=1}^{s} \alpha_j F(V_j) 
    +\sum_{l=1}^{p} \sum_{j=1}^{s} 
     G(V_j) \beta_{j,l}\theta_l \;\;,
\end{eqnarray}
where $v_n\approx v(t_n)$ and $v_{n+1}\approx v(t_{n+1})$ denote the numerical solutions at times $t_n$ and $t_{n+1}=t_n+\Delta t$, respectively, $\Delta t$ is the time step size, $p$ denotes the highest order of iterated Stratonovich integrals used in the scheme (here we consider only $p=1$ or $p=2$), $\theta_l$ is an $M$-dimensional Gaussian random vector, and $G(V_j)\theta_l$ denotes the product of the diffusion matrix $G$ and the vector $\theta_l$. For schemes with $p=1$, each component of $\theta_1$ is generated as an independent Gaussian random variable with zero mean and variance equal to $\Delta t$. For schemes with $p=2$, we have $\theta_1=R_1\sqrt{dt}$, $\theta_2=0.5\sqrt{dt}(R_1+R_2/\sqrt{3})$, where $R_1$ and $R_2$ are two independent $M$-dimensional Gaussian random vectors with independent components of zero mean and unit variance \cite{holm2018stochastic}. The coefficients $(a_{ij}, b_{ij}, \alpha_j, \beta_j)$ of the scheme are usually compactly written in the form of a tableau  \cite{burrage1996high}
\begin{eqnarray}
    \begin{matrix}
          a_{ij} & b_{ij,1} & b_{ij,2} & \ldots & b_{ij,p} \\
         \hline
          \alpha_i & \beta_{j,1} & \beta_{j,2} & \ldots & \beta_{j,p} 
    \end{matrix}\;\;.
\end{eqnarray}
The tableau reduces to the well-known Butcher tableau if the diffusion matrix is zero ($G=0$). If for $j\geq i$ we have that $a_{ij}=0$ and $b_{ij,l}=0$, then the numerical scheme is explicit; otherwise it is implicit. Stochastic Runge-Kutta methods for It\^o SDEs can be defined in a similar fashion.

The following schemes can be implemented in the integrator for the It\^o form.
\begin{enumerate}
    \item The Euler-Maruyama scheme
    \begin{eqnarray}
        \begin{matrix}
            \begin{matrix}
                & \begin{bmatrix} 0 \end{bmatrix}  &  \\
            \end{matrix}
             \begin{matrix}
                & \begin{bmatrix} 0 \end{bmatrix}  &  \\
            \end{matrix}
            \\\hline 
            \begin{matrix}
                & \begin{bmatrix} 1 \end{bmatrix}  &  \\
            \end{matrix}
             \begin{matrix}
                & \begin{bmatrix} 1  \end{bmatrix}  &  \\
            \end{matrix}
        \end{matrix}\;\;.
    \end{eqnarray}
    \item The splitting scheme can be written formally as follows \cite{strang1968construction,slaby2017combining}
    \begin{eqnarray}
        \begin{matrix}
            \begin{matrix}
                & \begin{bmatrix} 0 & 0 & 0 & 0 \\
                                  0.5 & 0 & 0 & 0 \\
                                  0 & 0 & 0 & 0 \\
                                  0 & 0 & 0.5 & 0
                  \end{bmatrix}  &  \\
            \end{matrix}
             \begin{matrix}
                & \begin{bmatrix} 0 & 0 & 0 & 0 \\
                                  0 & 0 & 0 & 0 \\
                                  0 & 1 & 0 & 0 \\
                                  0 & 0 & 0 & 0
                \end{bmatrix}  &  \\
            \end{matrix}
            \\\hline 
            \begin{matrix}
                & \begin{bmatrix} 0 & 0 & 0 & 1 \end{bmatrix}  &  \\
            \end{matrix}
             \begin{matrix}
                & \begin{bmatrix} 0 & 0 & 0 & 1  \end{bmatrix}  &  \\
            \end{matrix}
        \end{matrix}\;\;.
    \end{eqnarray}
\end{enumerate}

The following schemes can be implemented in the integrator for the Stratonovich form. 
\begin{enumerate}
    \item The Heun scheme (\cite{BurrageTian2004})
    \begin{eqnarray}
        \begin{matrix}
            \begin{matrix}
                & \begin{bmatrix} 0 & 0 \\
                                  1 & 0\end{bmatrix}  &  \\
            \end{matrix}
             \begin{matrix}
                & \begin{bmatrix} 0 & 0 \\
                                  1 & 0 \end{bmatrix}  &  \\
            \end{matrix}
            \\\hline 
            \begin{matrix}
                & \begin{bmatrix} \frac{1}{2} & \frac{1}{2} \end{bmatrix}  &  \\
            \end{matrix}
             \begin{matrix}
                & \begin{bmatrix}  \frac{1}{2}  &  \frac{1}{2}   \end{bmatrix}  &  \\
            \end{matrix}
        \end{matrix}\;\;.
    \end{eqnarray}
    \item A two-stage scheme (referred to as ``PL''(Platen)) is written with $s=2$, $p=1$ (Eq. 27 of \cite{burrage1996high}) as
    \begin{eqnarray}
        \begin{matrix}
            \begin{matrix}
                & \begin{bmatrix} 0 & 0 \\ 1 & 0 \end{bmatrix}  &  \\
            \end{matrix}
             \begin{matrix}
                & \begin{bmatrix} 0 & 0 \\ 1 & 0 \end{bmatrix}  &  \\
            \end{matrix}
            \\\hline 
            \begin{matrix}
                & \begin{bmatrix} 1 & 0  \end{bmatrix}  &  \\
            \end{matrix}
             \begin{matrix}
                & \begin{bmatrix} \frac{1}{2} & \frac{1}{2}  \end{bmatrix}  &  \\
            \end{matrix}
        \end{matrix}\;\;.
    \end{eqnarray}
    
    \item A four-stage scheme (``E1") with $s=4$, $p=2$ (\cite{burrage2000order}) as
    \begin{eqnarray}{
        \begin{matrix}
            \begin{matrix}
                & \begin{bmatrix} 0 & 0 & 0 &0 \\ \frac{2}{3} & 0 & 0 & 0 \\ 
                \frac{3}{2} & -\frac{1}{3} & 0  & 0 \\ \frac{7}{6} & 0 & 0 & 0 \end{bmatrix}  &  \\
            \end{matrix}
             \begin{matrix}
                & \begin{bmatrix} 
                0  &  0  & 0  & 0 \\
                \frac{2}{3} & 0 & 0 & 0\\
                \frac{1}{2}  & \frac{1}{6} & 0 & 0 \\
                -\frac{1}{2}  & 0    & \frac{1}{2} & 0 \end{bmatrix}  &  \\
            \end{matrix}
            \begin{bmatrix}
                0 &0& 0& 0 \\
                0 & 0 & 0 & 0 \\
                -\frac{2}{3} & 0 & 0 & 0\\
                 \frac{1}{6}& \frac{1}{2} & 0 & 0
            \end{bmatrix}
            \vspace{.1cm}
            \\\hline 
            \begin{matrix}
                & \begin{bmatrix} \frac{1}{4} & \frac{3}{4}& -\frac{3}{4}&\frac{3}{4}  \end{bmatrix}  &  \\
            \end{matrix}
             \begin{matrix}
                & \begin{bmatrix} -\frac{1}{2} & \frac{3}{2} & -\frac{3}{4} & \frac{3}{4}  \end{bmatrix}   
            \end{matrix}
             \begin{matrix}
                & \begin{bmatrix}\frac{3}{2} & -\frac{3}{2} & 0 & 0 \end{bmatrix}   
            \end{matrix}            
        \end{matrix}\;\;.} 
    \end{eqnarray}    
\end{enumerate}
More high-order schemes can be found in \cite{burrage1996high,burrage2000order}. We have also listed several other schemes in \ref{app:schemes}

\section{Higher order integrators for the collision operator}
\label{sec:results}

\subsection{Numerical tests using analytical solutions}
\subsubsection{Strong and weak convergence of different schemes for 1D1W}
\label{subsubsec:1d1w}
\begin{figure}
\centering
\includegraphics[width=0.42\textwidth]{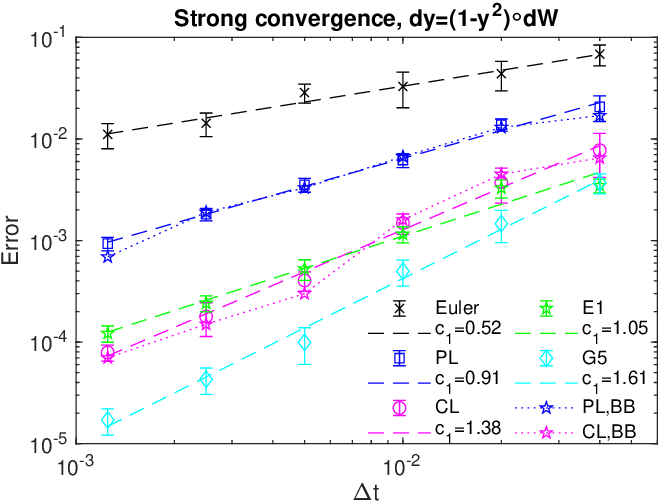}
\includegraphics[width=0.42\textwidth]{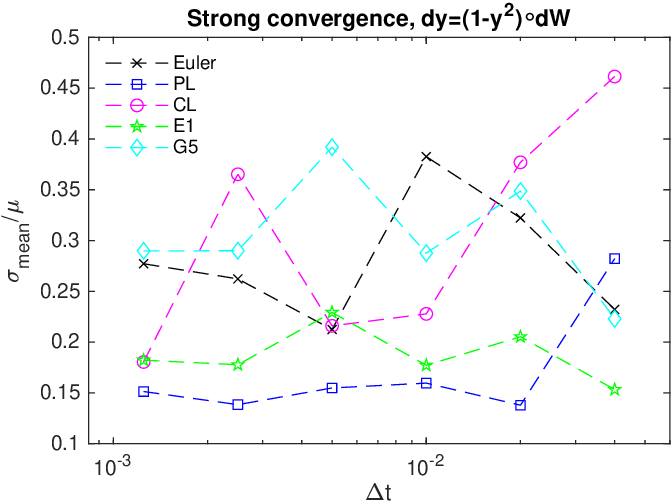}
\includegraphics[width=0.42\textwidth]{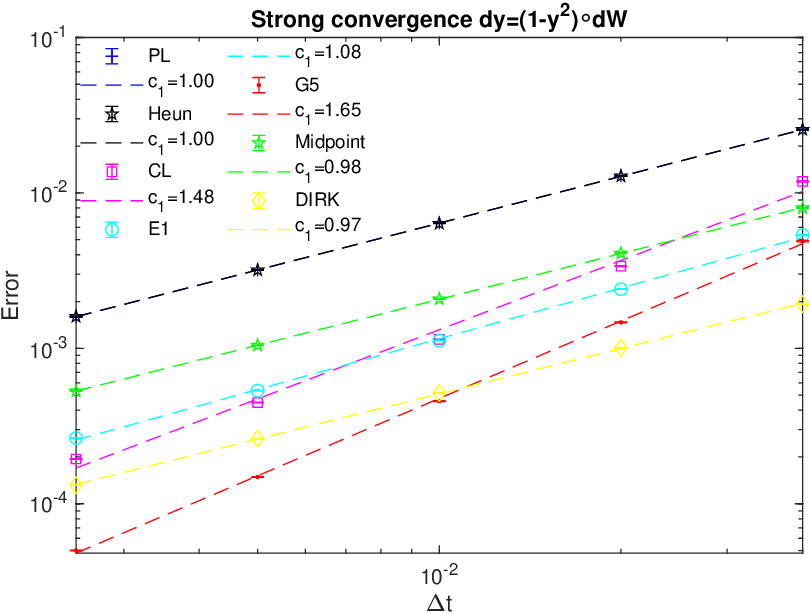}
\includegraphics[width=0.42\textwidth]{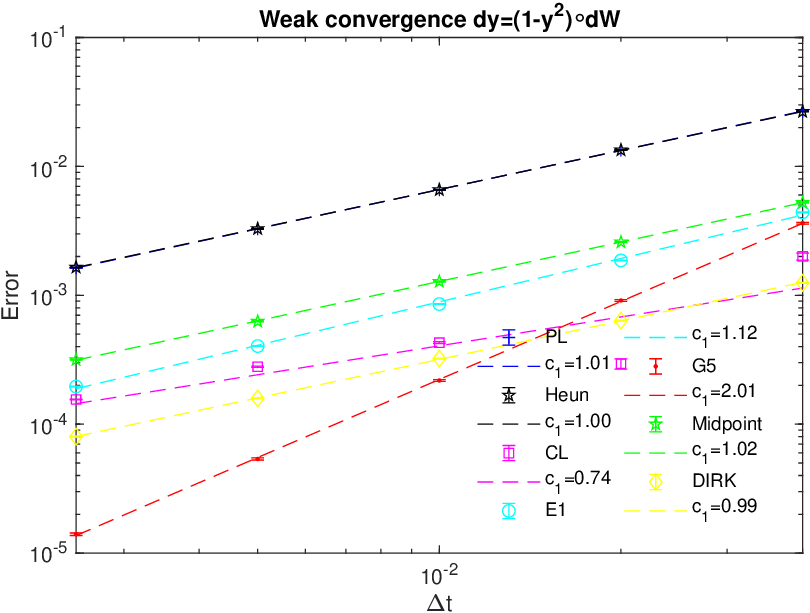}
\caption{\label{fig:compare_sde_scheme_error} \emph{Top left:} The strong errors of the numerical solution at the last step for the Euler-Maruyama scheme, the 2-stage PL scheme, and the 4-stage CL scheme using 25 sample paths are depicted. \emph{Top right:} The standard deviation of the mean of the errors. \emph{Bottom:} The same cases are run using 50000 sample paths. The strong (\emph{Left}) and weak (\emph{Right}) errors are depicted.
}
\end{figure}

As a basic test of different schemes, we consider a one-dimensional problem driven by a one-dimensional Wiener process, in the It\^o form described by the SDE (1D1W Problem 1 of \cite{burrage1996high}),
\begin{eqnarray}
    dy=-a^2y(1-y^2)dt+a(1-y^2)dW \;\;, \;\; y(0)=y_0\;\;,\;\; t\in[0,1],
\end{eqnarray}
which is equivalent to the Stratonovich SDE
\begin{eqnarray}
    dy=a(1-y^2)\circ dW \;\;, \;\; y(0)=y_0\;\;,\;\; t\in[0,1],
\end{eqnarray}
and whose solution is given analytically by
\begin{eqnarray}
    y_{theory}(t)=\tanh[aW(t)+\text{arctanh}(y_0)].
\end{eqnarray}
As the initial condition, we choose $y_0=0.5$. Other parameters are the same as those in \cite{burrage1996high}, namely, $a=1$, $\Delta t=0.00125$, $0.0025$, $0.005$, $0.01$, $0.02$, $0.04$. For each scheme, 25 trajectories (also referred to as particles or samples) are launched and the averaged error is calculated at $t_{end}=1$ when the results from the solver developed in this work are compared with the previous work \cite{burrage1996high}.  The global error $\epsilon_{y,i}$ is calculated for each trajectory $y_i(t)$ first. Then the mean error $\bar\epsilon_y$ (the so-called strong error) and the standard deviation of the mean (which serves as a measure of the Monte Carlo error occurring due to the finite number of samples in the computations) are calculated for all particles. Namely,
\begin{eqnarray}
\label{eq:eps1d1w}
    \epsilon_{y,i}&=&|y_i(t_{end})-y_{theory,i}(t_{end})|\;\;, \\
    \label{eq:mean_error}
    \bar\epsilon_y &=&\frac{1}{N_p}\sum_{i=1}^{N_p} \epsilon_{y,i}\;\;, \\
    \label{eq:standard_deviation_of_mean}
    \sigma_{mean}&=&\sqrt{\frac{1}{N_p(N_p-1)} \sum_{i=1}^{N_p}(\epsilon_{y,i}-\bar\epsilon_y)^2} \;\;.
\end{eqnarray}
As shown in the top left frame of Fig. \ref{fig:compare_sde_scheme_error}, we calculated the strong error for different schemes for different values of time step size $\Delta t$. The half-height of the error bar is $\sigma_{mean}$. The previous results for the PL and CL schemes in \cite{burrage1996high} are also shown for benchmarking. Reasonable agreement can be seen, while the discrepancy is due to the small ensemble size (we used only 25 particles to mimic the experiment in \cite{burrage1996high}). The right frame shows the standard derivation of the mean of the errors divided by the mean error, indicating that the half-height of the error bar in the left frame is much smaller (by a factor of $0.15\sim0.45$) than the mean error. The higher-order schemes (4-stage CL, E1, and 5-stage G5) reduce the relative error significantly than the Euler-Maruyama scheme. The convergence order is calculated by the least-square fitting with two free parameters $c_0, c_1$,
\begin{eqnarray}
    \log\epsilon = c_0+c_1\log dt \;\;,
\end{eqnarray}
where $c_1$ is the convergence order shown in Fig. \ref{fig:compare_sde_scheme_error}. The convergence order is calculated in a similar way for the other cases in the following sections. Theoretically, the Euler, PL, E1, CL, and G5 schemes have a strong order of convergence of $0.5$, $1.0$, $1.0$, $1.5$, and $1.5$, respectively \cite{burrage1996high,burrage2000order}. The strong convergence order $c_1=0.52$ observed in the numerical experiment for the Euler-Maruyama scheme is close to the theoretical value $0.5$. Those for PL, E1, CL, and G5, $c_1=0.91, 1.05, 1.38,  1.61$,  are slightly different from the expected theoretical values $c_{1,theory}=1, 1, 1.5, 1.5$ as shown in the top-left frame. Nevertheless, the high-order schemes show their advantages in reducing errors. For $\Delta t=0.00125$, the relative errors of PL and CL schemes are lower than $10^{-1}$ and $10^{-2}$ of that of the Euler-Maruyama scheme respectively. 

To investigate the effect of the ensemble size and to reduce the Monte Carlo error, the same cases were studied with $50000$ particles/samples. The similar strong convergence orders of the PL (Platen), E1, GL, and G5 schemes are obtained as shown in the bottom left frame. In addition, the error of the Heun scheme is also shown, which is higher than that of other schemes except that of PL (and also Euler-Maruyama). The Midpoint and DIRK (diagonally implicit Runge–Kutta) schemes are implicit schemes that have been studied previously \cite{tyranowski2021stochastic} and they (especially DIRK) give better precision than the explicit high-order schemes. 
To assess the weak convergence of the considered integrators, we also studied how well the numerical expected value $\mathbb{E}(y^2(t_{end}))$ approximates the theoretical expected value $\mathbb{E}(y^2_{theory}(t_{end}))$. The weak error $\bar \eta_y$ and the corresponding standard deviation of the mean $\sigma_{weak}$ are defined by the formulas
\begin{eqnarray}
    \label{eq:delta1D1W}
    \delta_y &=&\frac{1}{N_p}\sum_{i=1}^{N_p} \Big( y^2_i(t_{end})-y^2_{theory,i}(t_{end}) \Big)\;\;, \\
    \label{eq:weak_error}
    \bar\eta_y &=&|\delta_y|\;\;, \\
    \label{eq:standard_deviation_of_mean for the weak error}
    \sigma_{weak}&=&\sqrt{\frac{1}{N_p(N_p-1)} \sum_{i=1}^{N_p}\Big(y^2_i(t_{end})-y^2_{theory,i}(t_{end})-\delta_y\Big)^2} \;\;.
\end{eqnarray}
The weak convergence plots are depicted in the bottom right frame of Fig.~\ref{fig:compare_sde_scheme_error}.

\subsubsection{Kubo oscillator}
\label{subsec:kubo}
The Kubo oscillator is the two-dimensional Stratonovich SDE
\begin{eqnarray}
    dq&=& +pdt + \gamma p\circ dW \;\;, \\
    dp&=& -qdt - \gamma q\circ dW \;\;,
\end{eqnarray}
driven by the standard one-dimensional Wiener process $W(t)$, where $p,q$ are the oscillator's phase space coordinates, and $\gamma$ is the noise intensity. The analytical solution is given by
\begin{eqnarray}
    q_{theory}(t)&=&q_0\cos[t+\gamma W(t)] + p_0 \sin[t+\gamma W(t)] \;\;, \\
    p_{theory}(t)&=&p_0\cos[t+\gamma W(t)] - q_0 \sin[t+\gamma W(t)] \;\;.
\end{eqnarray}
The energy conservation of the Kubo oscillator is readily obtained,
\begin{eqnarray}
    E=q^2(t)+p^2(t)=q_0^2+p_0^2\;\;. 
\end{eqnarray}
Considering the analytical solution and the energy conservation property of the Kubo oscillator, we show strong convergence (accuracy of the trajectory) and weak convergence (energy conservation) of various schemes. 

\begin{figure}
\centering
\includegraphics[width=0.3\textwidth]{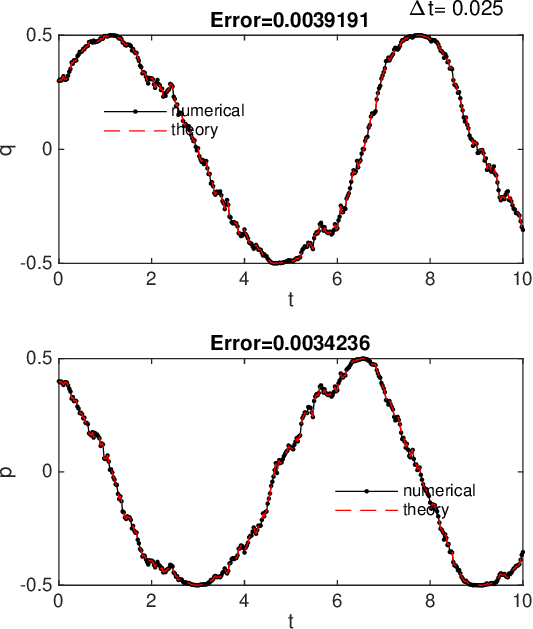}
\includegraphics[width=0.3\textwidth]{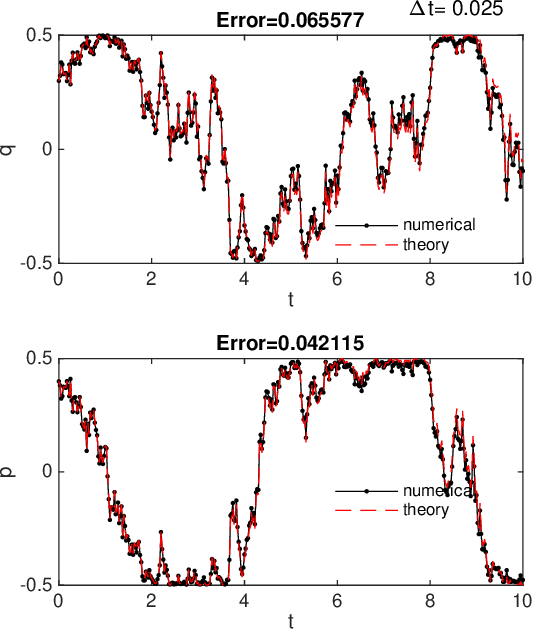}
\caption{\label{fig:t1d_Kubo} The time evolution of the Kubo oscillator for $\gamma=0.25$ (left) and $1$ (right).  }
\end{figure}

\begin{figure}
\centering
\includegraphics[width=0.98\textwidth]{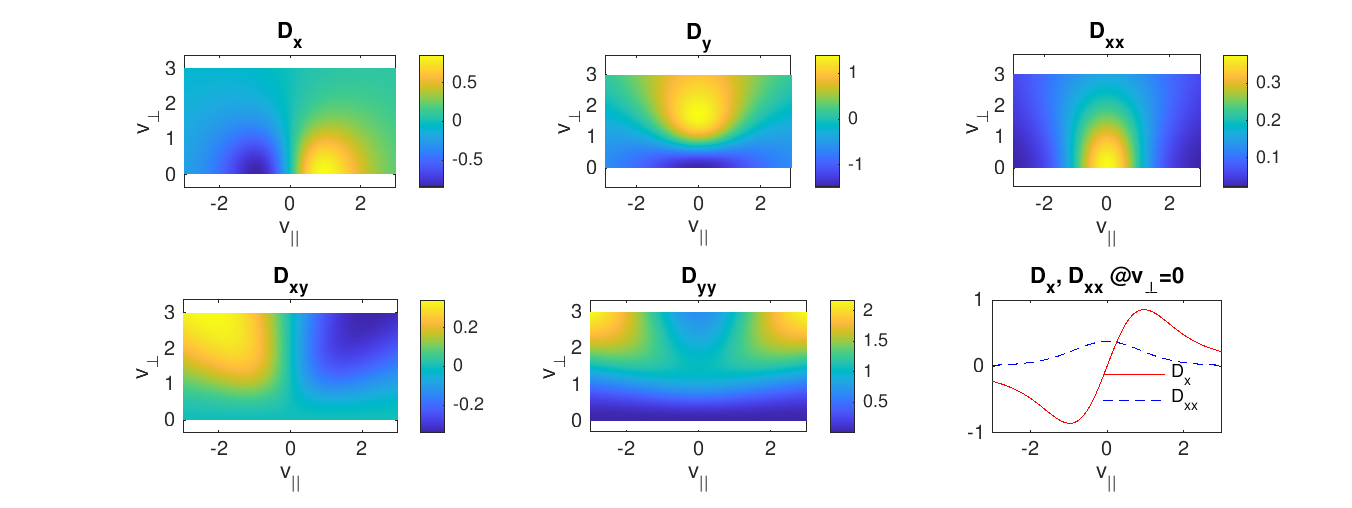}
\caption{\label{fig:Dcoef2dpde} The structure of the drag and diffusion coefficients in Eq. \ref{eq:langevin_dfdt}. }
\end{figure}

In the numerical studies, the initial condition is chosen as 
\begin{eqnarray}
\label{eq:kubo_initial}
    q_0=0.3 \;\;,\;\;
    p_0=0.4\;\;.
\end{eqnarray}
The coefficient $\gamma$ gives the magnitude of the noise process. Two cases with $\gamma=0.25$ and $\gamma=1$ are shown in Fig. \ref{fig:t1d_Kubo}. When the diffusion part is small, namely, $\gamma=0.25$, the time evolution is similar to that of the harmonic oscillator, as shown in the left frame. When the diffusion part is larger, namely, $\gamma=1$, the time evolution becomes more stochastic as shown in the right frame. Since in the collision operator the drift and diffusion parts are of the same magnitude, as will be shown in Fig. \ref{fig:Dcoef2dpde}, we choose $\gamma=1$ in this section so that it is more relevant to the studies of the collision operators. The calculation of errors is the same as that in Sec. \ref{subsubsec:1d1w} except the absolute value in Eq. \ref{eq:eps1d1w}
\begin{eqnarray}
\label{eq:eps2d1w}
    \epsilon_{i}&=&\sqrt{[q_i(t_{end})-q_{i,theory}(t_{end})]^2+[p_i(t_{end})-p_{i,theory}(t_{end})]^2}\;\;,
\end{eqnarray}
where `$i$' is the particle/trajectory index.

\begin{figure}
\centering
\includegraphics[width=0.425\textwidth]{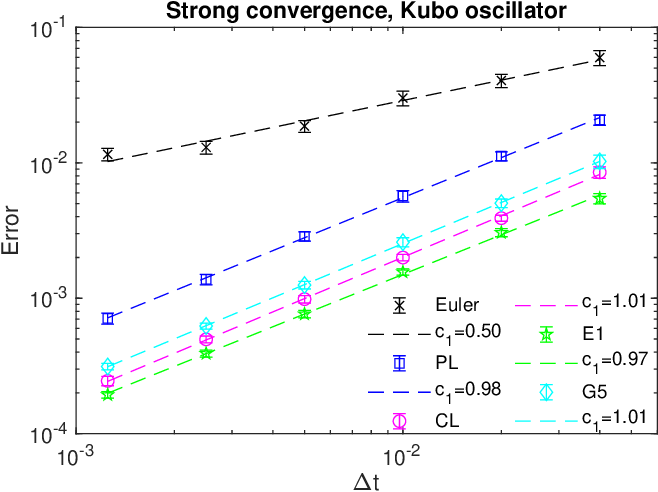}
\includegraphics[width=0.425\textwidth]{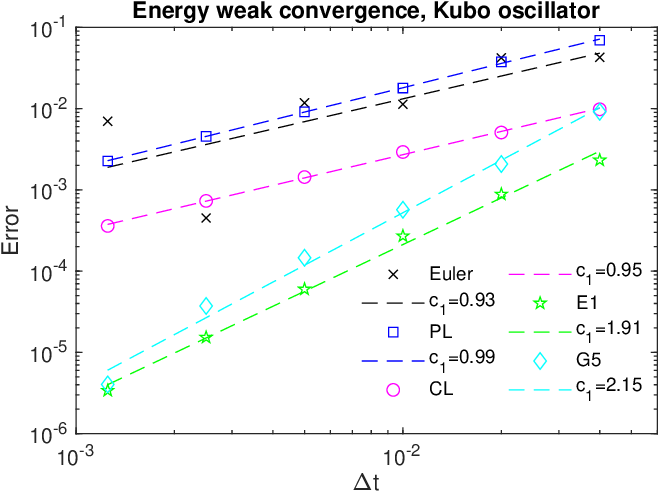}
\includegraphics[width=0.425\textwidth]{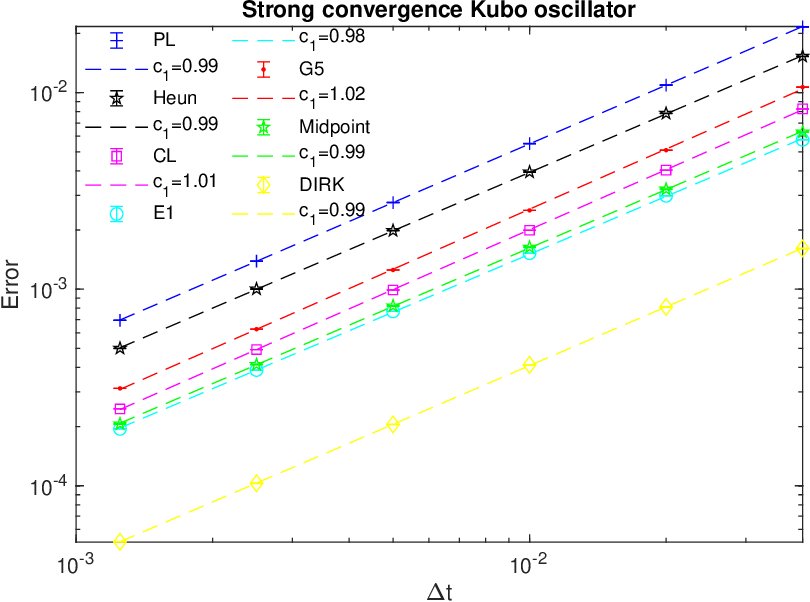}
\includegraphics[width=0.425\textwidth]{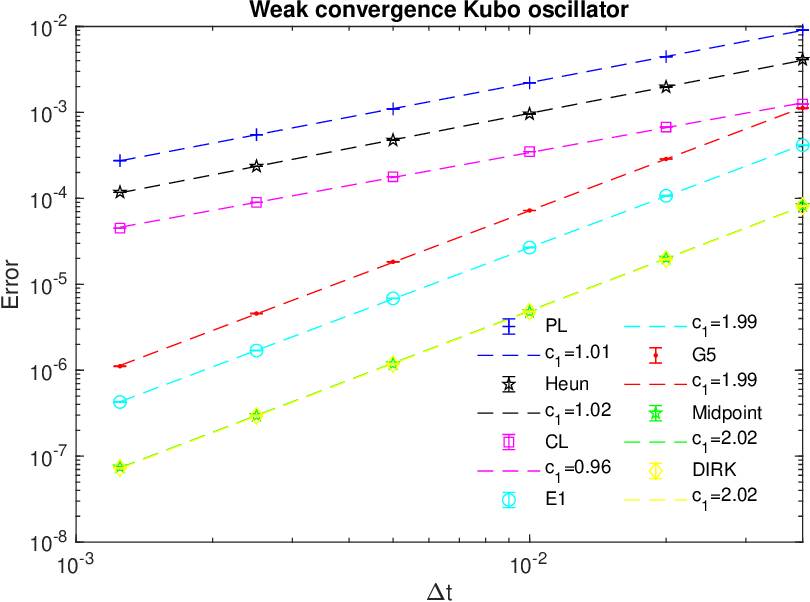}
\caption{\label{fig:compare_Kubo_converge} The strong error at the end of the simulation (top left), the relative error of the total energy (top right) for the Kubo oscillator with 100 particles.  The same cases of the Kubo oscillator are run using 50000 particles as shown at the bottom, with another four schemes included for comparison.}
\end{figure}

The weak and strong convergence properties are studied by launching $100$ (top frames) and $50000$ (bottom frames) trajectories, with $\gamma=1$, $t_{end}=1$, $\Delta t=0.00125$, $0.0025$, $0.005$, $0.01$, $0.02$, $0.04$ and the initial condition in Eq. \ref{eq:kubo_initial}.  The strong convergence is shown in the left frame of Fig. \ref{fig:compare_Kubo_converge}. The 2-stage PL, 4-stage CL, E1, and the 5-stage G5 show strong convergence of order $\sim1$, and the magnitude of the error is also significantly lower than the Euler-Maruyama scheme. For $\Delta t=0.00125$, the error of the PL scheme is lower than $10\%$ of that of the Euler-Maruyama scheme, and the CL, E1, and G5 relative errors are lower than $2\%$ of that of the Euler-Maruyama scheme.
The weak energy convergence is shown in the right frame of Fig. \ref{fig:compare_Kubo_converge}. The Euler-Maruyama scheme shows a weak convergence of order $\sim1$, the same as the PL and CL schemes. The PL scheme has an even higher error than the Euler-Maruyama scheme. For cases with $100$ trajectories, the relative error of the CL scheme is lower than $1/5$ of that of the Euler-Maruyama. For  E1 and G5, the convergence orders are $1.9$ and $2.1$ respectively, and the magnitude of the error at $\Delta t=0.00125$ is lower than the Euler-Maruyama by a factor of $\sim1/300$. 

The bottom frames depict also the numerical results for the implicit midpoint and DIRK schemes, as well as for the weakly convergent RS1 and RS2 schemes \cite{rossler2007second}. The error curves of RS1 and RS2 almost overlap with each other, indicating similar advantages in the precision in the weak sense. While the four schemes (Midpoint, DIRK, RS1, and RS2) have their good properties, the implementation of the former two is more complicated due to the implicit treatment, and the latter two do not guarantee strong convergence. Thus, in the collision studies, we focus on the explicit high-order schemes and leave the applications of the implicit and weakly convergent schemes to our future work. 
Overall, among the explicit high-order schemes, the E1 scheme shows better strong and weak convergence properties than the Euler-Maruyama, PL, CL, and G5 schemes. In the following studies of collision operators, we only consider the Euler-Maruyama, PL, and E1 schemes, the former two of which are considered as reference schemes for the E1 scheme.

\subsection{The high-order stochastic integrator for the collision operators}
\subsubsection{Features of the Rosenbluth-Trubnikov collision operator}
For collision operator with Maxwellian background, the drag and the diffusion coefficients in Eqs. \ref{eq:langevin_drag}--\ref{eq:langevin_diffusion} are shown in Fig. \ref{fig:Dcoef2dpde}. The general form of the diffusion and drag coefficients derived from the Rosenbluth-Trubnikov potentials are not analytically tractable. Both coefficients are fully non-uniform in the velocity space and are unknown a priori.  Thus the corresponding stochastic equation is non-commutative.

Compared with the Kubo oscillator discussed in Section \ref{subsec:kubo}, the collision operator problem is different in several aspects as follows.
\begin{enumerate}
    \item The analytical solution of the Langevin equation corresponding to the collision operator can be hardly given. On the other hand, conservation is one key issue for physics studies of our interest. Thus the strong convergence is not studied in this work for the collisions. 
    \item Generally, for the whole multiple-species particle system, the conservation of energy and momentum is satisfied. In our case, since the test particle with a Maxwellian distribution is given, if the temperature of the test particle is the same as the background particles that are used for the calculation of the Rosenbluth-Trubnikov potentials, the energy conservation and momentum conservation are satisfied.  
    \item Generally, for each particle, the energy and the momentum are not conserved since an individual particle can migrate from one location to another in the velocity space. 
\end{enumerate}
In numerical studies, numerous particles are initialized with a given analytical distribution, and the collisions are applied. 
Since the particles represent the distribution, even if the analytical solution of $f$ is known in the presence of collisions, the calculated $f$ from markers and the total energy of all particles can deviate from the analytical solution due to the noise, which causes the Monte-Carlo error. As a result, to evaluate the performance of the high-order stochastic integrator, the marker number should be large enough so that the Monte-Carlo error is lower than the error due to the time integrator. For the purpose of this experiment, we adopt the number $\epsilon_{acc}\equiv 1/\sqrt{N_p}$ as an acceptable error threshold, that is, we only analyze the error of the total energy if it is higher than $\epsilon_{acc}$.

The boundary condition at $y=0$ is applied to particles. In the original partial differential equation, there is no flux flowing to the $y<0$ region. In solving the SDE of each particle, in one step, if a particle's $y$ becomes negative, then this step is canceled and the random number is re-generated until $y$ ends up with a positive value (``step-back-re-try''). Other treatments are possible, for example, A) set $y$ to zero; B) set $y$ to its absolute value, and C) step back without re-trying, if negative $y$ appears. However, in the following numerical studies, we adopt the ``step-back-re-try'' scheme if $y<0$ since other treatments do not give better conservation of energy in our studies. 

\subsubsection{The Rosenbluth-Trubnikov potentials from the mixed-MC-FEM solver}
\begin{figure}
\centering
\includegraphics[width=0.45\textwidth]{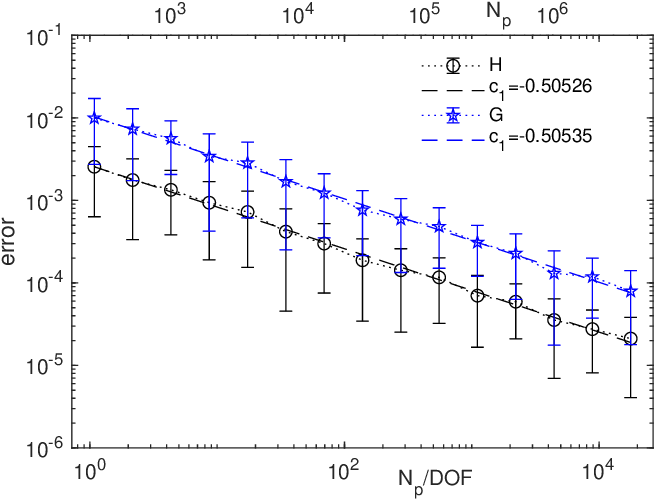}
\caption{\label{fig:hg_converge} The error of the Rosenbluth-Trubnikov potentials $h$ and $g$ from the mixed-MC-FEM solver. The bottom $x$ and top $x$ labels are the marker number per degree of freedom of the finite element solver and the total marker number, respectively. }
\end{figure}

The Rosenbluth-Trubnikov potentials are obtained by solving Eqs. \ref{eq:elliptic_h}, \ref{eq:elliptic_g} with the boundary conditions given by Eqs. \ref{eq:neumann_h}, \ref{eq:neumann_g}, \ref{eq:h_pic}, \ref{eq:g_pic}. The 2-dimensional cubic splines are adopted in the finite element scheme. The error of $h$ and $g$ is shown in Fig. \ref{fig:hg_converge}. The grid numbers in $(v_\|,v_\perp)$ are $N_{v_\|}=21$, $N_{v_\perp}=11$. The error is calculated by making use of the analytical solution of $h$ and $g$ for Maxwellian background distribution in Eqs. \ref{eq:h_theory} and \ref{eq:g_theory}, denoted as $h_{theory}$ and $g_{theory}$,
\begin{eqnarray}
    \epsilon_A = \sqrt{\frac{\sum_{i,j}(A_{i,j}-A_{i,j,theory})^2}{\sum_{i,j}A_{i,j,theory}^2}}\;\;,
\end{eqnarray}
where $A=h,g$ and $i,j$ denote the grid indices along $v_\|$ and $v_\perp$, respectively.
The error is shown in Fig. \ref{fig:hg_converge}. The convergence order is $\sim0.5$. Since the cubic spline finite element gives high enough accuracy, the main error is from the marker noise. In this case, using $1000$ markers per degree of freedom of the finite element solver gives reasonably low errors ($\epsilon_A<10^{-3}$) for both $h$ and $g$. 

While the study of the mixed-MC-FEM solver is for the completeness of the collision model in particle simulations, one of our focuses is the application of the high-order SDE schemes.
Thus, in the rest part of this work, the background distribution is chosen as the Maxwellian distribution, and the Rosenbluth-Trubnikov potentials are obtained analytically so that the performance of the high-order SDE schemes can be evaluated separately.

\subsubsection{Maintainance of Maxwellian test particles}
\label{subsubsec: Maintainance of Maxwellian test particles}
\begin{figure}
\centering
\includegraphics[width=0.45\textwidth]{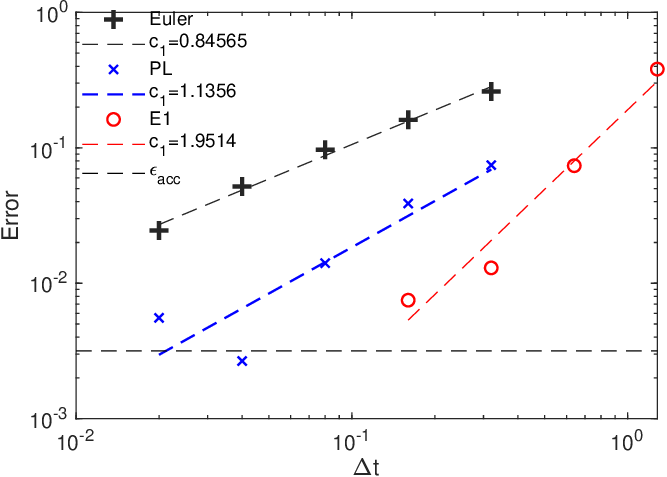}
\includegraphics[width=0.455\textwidth]{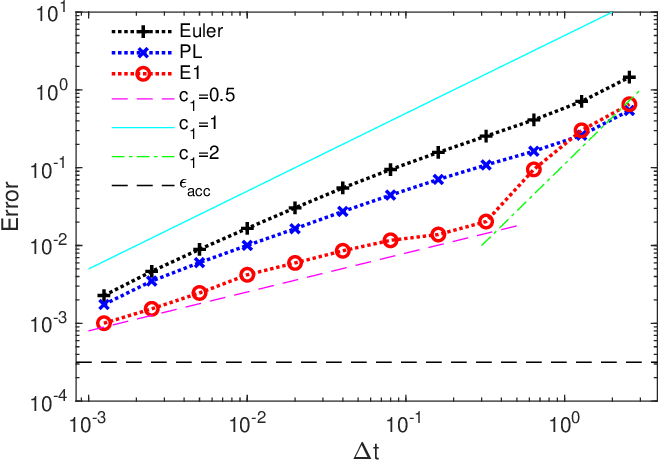}
\caption{\label{fig:collision_weak_convergence} The relative error of the total particle kinetic energy for different values of the time step size $\Delta t$ and for different schemes. Left: $10^5$ markers; right: $10^7$ markers. }
\end{figure}

We choose the Maxwellian distribution for the test particles with the same temperature as the background Maxwellian particles. 
The expected theoretical solution is the maintenance of the Maxwellian distribution function after the collisions are applied.  
The energy conservation is checked for various schemes.
The Monte-Carlo noise due to finite particle number thus gives a constraint of the achievable lowest relative error of the total particle energy. However, the high-order schemes of the SDE are still expected to improve the overall computation accuracy when the time integrator-induced error is higher than that due to the Monte-Carlo noise. As shown in Fig. \ref{fig:collision_weak_convergence}, the relative error of total particle energy is shown for different values of time step size $\Delta t$. In these cases, we have used $N_p=10^5$ (the left frame) and $10^7$ (the right frame) markers and thus the acceptable error thresholds $\epsilon_{acc}$ are $\sim 1/\sqrt{N_p}\approx 3.3\times 10^{-3}$, $3.3\times 10^{-4}$, respectively, as indicated by the horizontal black dashed line. 
In the left frame, the Euler-Maruyama scheme leads to a higher relative error than the PL scheme by $\sim8$ times for $\Delta t=0.08$. For higher-order schemes, the acceptable error threshold $\epsilon_{acc}$ is reached when $\Delta t$ is small, and thus the relative error at low $\Delta t$ is mainly determined by the particle number. The weak convergence of the 4-stage scheme is better than the 2-stage scheme and the Euler-Maruyama scheme. In other words, to achieve the same accuracy, the allowed maximum time step size $\Delta t$ can be larger if using higher-order schemes than using the Euler-Maruyama scheme. 
In the right frame, as the larger marker number ($N_p=10^7$) is used, the relative errors for smaller time step sizes are also studied. For the E1 scheme, the convergence order becomes smaller than $2$ for $\Delta t<0.16$ and is between $0.5$ and $1$. At $\Delta t=0.00125$, the relative error of E1 is lower than the Euler-Maruyama scheme by a factor of $\sim1/2$, while at $\Delta t=0.16$, by a factor of $\sim1/11$, which indicates that the E1 scheme's advantage is more significant for moderate to large $\Delta t$ values. 

\subsubsection{Thermalization of a test particle distribution}
The thermalization of the test Maxwellian particle $a$ due to collisions with fixed Maxwellian field species $b$  \cite{trubnikov1965particle} is chosen as the test case in this section, as is also studied previously using the finite volume solver \cite{xiong2008high}. The distribution of the test particle is assumed to be Maxwellian, and the theoretical rate of the temperature change of the test particle is given by 
\begin{eqnarray}
    \frac{dT_a}{dt} = -\frac{8m_a}{3m_b\sqrt{\pi}} \frac{T_a-T_b}{\tau^{a\backslash  b}(T_a+\frac{m_a}{m_b}T_b)} \;\;,
\end{eqnarray}
where $\tau^{a\backslash b}$ is the basic relaxation time defined by $\tau^{a\backslash b}=\sqrt{m_a}\varepsilon^{3/2}/(\pi \sqrt{2}e_a^2e_b^2\Lambda_c n_b)$. 
The steady-state solution for which $dT_a/dt=0$ can be readily obtained as $T_a=T_b$. In our numerical test, we assume the temperature of the background particle is fixed, namely, $T_b(t)=T_b(t=0)$, yielding the steady-state solution $T_a(t\gg\tau^{a\backslash b})=T_b$.

\begin{figure}
\centering
\includegraphics[width=0.45\textwidth]{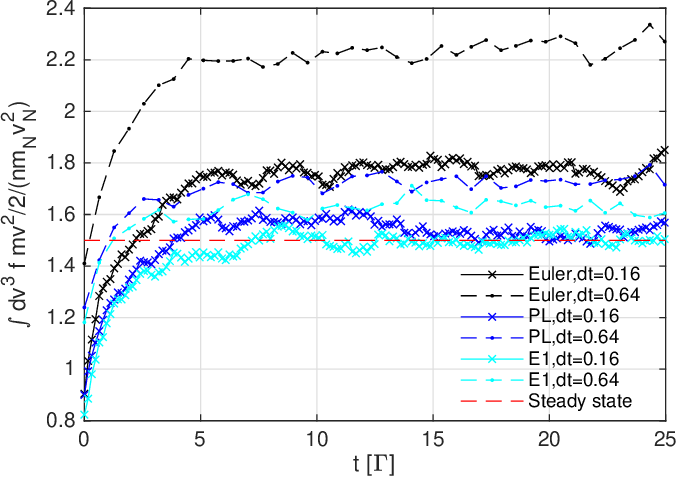}
\includegraphics[width=0.45\textwidth]{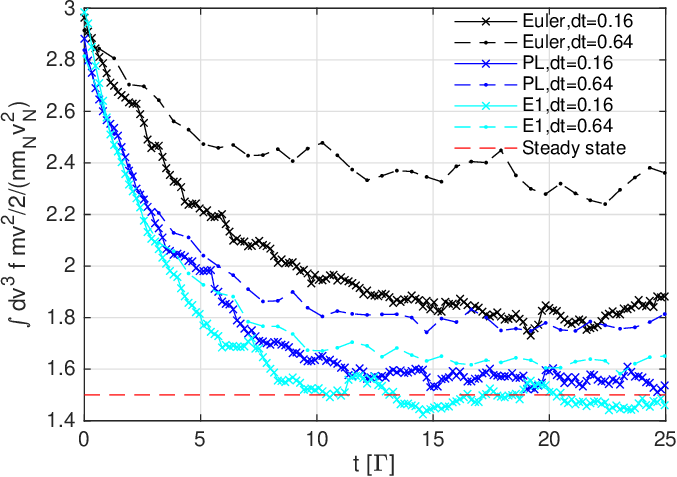}
\caption{\label{fig:thermalize_isotropic} The mean energy evolution of the thermalization of the test particles with the initial condition $T_a/T_N=0.5$ (left) and $T_a/T_N=2$ (right).}
\end{figure}

Three schemes are compared for different time step sizes, as shown in Fig. \ref{fig:thermalize_isotropic}. The time step size has to be sufficiently small to produce the correct final temperature of the test particle distribution. The high-order scheme (the 4-stage E1 scheme) produces the final temperature precisely for $\Delta t=0.16$. Still, the error using the Euler-Maruyama scheme is significant and thus smaller $\Delta t$ is needed to achieve the same accuracy as that of the high-order E1 scheme. 

\subsubsection{Relaxation of anisotropic distribution to isotropic Maxwellian}
In this section, we demonstrate that the implemented  collision operator can relax an anisotropic distribution to an isotropic Maxwellian distribution. The relaxation of the distribution with loss cone cavity to isotropic Maxwellian distribution is chosen in the following studies, as done previously in the finite-volume work \cite{xiong2008high}. The loss cone is defined in the velocity space $(\Lambda, E)$, where $\Lambda=v_\perp^2/v^2$, $E=v^2/2$. As the initial condition, the particles in the loss cone $\Lambda_0<\Lambda<\Lambda_1$ are eliminated, to mimic the physics process due to particle loss in the specific regime of the velocity space. 

\begin{figure}
\centering
\includegraphics[width=0.45\textwidth]{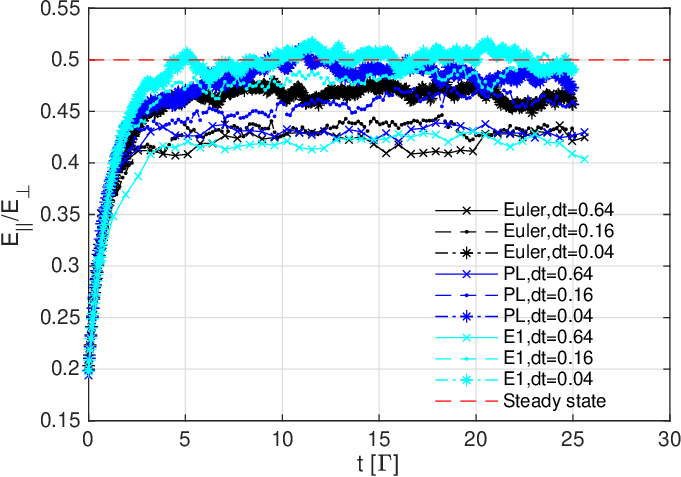}
\caption{\label{fig:loss_cone_relax} The relaxation of the distribution with a loss-cone cavity to the isotropic Maxwellian distribution. }
\end{figure}

The relaxation of the distribution with a loss-cone cavity to the isotropic Maxwellian distribution is shown in Fig. \ref{fig:loss_cone_relax}. As the initial condition, the particles with $0\le\Lambda\le 0.5$ are absent. The high-order scheme shows its advantages in reaching a more accurate final solution. For a large step size $\Delta t$, all schemes give the final state with significant error in $E_\|/E_\perp$. As $\Delta t$ decreases, $E_\|/E_\perp$ converges to the theoretical value for $\Delta t=0.04$ if using a high-order scheme  but for the  Euler-Maruyama scheme, a relative error of the order of $5\%$ in the total energy is produced in the steady state.

\section{Summary}
\label{sec:summary}
In this study, we have applied high-order schemes to stochastic differential equations, both in the Stratonovich and the It\^o forms, associated with the Fokker-Planck equation with the Rosenbluth-Trubnikov collision operator. We have formulated a mixed Monte Carlo-Finite Element Method for the calculation of the nonlinear collision operator for a general distribution function. We have implemented various high-order stochastic integration schemes and we have demonstrated their favorable properties, such as better accuracy and improved conservation of energy. More specifically, in the numerical simulations related to the Fokker-Planck equation with the Rosenbluth-Trubnikov collision operator with Maxwellian background particles we have observed that:
\begin{itemize}
    \item The  4-stage E1 scheme shows better weak energy convergence than the Euler-Maruyama and PL schemes. 
    \item The weak energy convergence order of the E1 scheme is between $1\sim2$, while the Euler-Maruyama scheme order is $\sim 1$ for moderate to large time step sizes ($\Delta t\ge 0.32$). For small time step sizes, the convergence order of the Euler-Maruyama, PL, and E1 schemes are  $0.5\sim1$. 
    \item The magnitude of the relative error of the E1 scheme is lower than the Euler-Maruyama scheme by a factor of $1/4$ at $\Delta t=0.01$, by a factor of $1/12$ at $\Delta t=0.32$. 
    \item The high-order E1 scheme produces reasonably accurate physics results of thermalization of isotropic and anisotropic particles at moderate or relatively large time step size $\Delta t=0.16$, $0.04$ respectively, while the Euler-Maruyama scheme gives significantly larger errors.
\end{itemize}
The application of these high-order stochastic integrators to neoclassical transport can make the simulation more efficient in particle simulations, specifically in the test particle collision part in the linearized collision operators \cite{satake2020benchmark}. In addition, it can be readily noticed that the collisionless process of the guiding centers' motion can be taken into account in the drift part of the stochastic differential equation using high-order schemes, offering the potential implementation of such advanced schemes in existing gyrokinetic particle codes.

Future work will focus on the study of fully nonlinear collision operators. Since the Rosenbluth-Trubnikov potentials are calculated using Maxwellian distribution when the high-order SDE schemes are applied, the more consistent way is to use the realistic distribution which can vary self-consistently. 
In addition, it also merits more effort to apply the collision operators using high-order schemes to physics studies such as neoclassical transport near the edge or/and in the steep density and temperature gradient regimes, as well as the collisional transport in the plasma edge in toroidally confined plasmas to extend the previous work \cite{lin1995gyrokinetic,rekhviashvili2023}.  

Another direction in which our work can be extended is the application of stochastic geometric and variational integrators. \emph{Geometric integration} of Hamiltonian systems has been thoroughly studied (\cite{HLWGeometric}) and geometric integrators have been shown to demonstrate superior performance in long-time simulations of such systems, compared to non-geometric methods. An important class of geometric integrators are \emph{variational integrators}. This type of numerical schemes is based on discrete variational principles and provides a natural framework for the discretization of Lagrangian systems (\cite{MarsdenWestVarInt,MarsdenPatrickShkoller,LeokZhang,OberBlobaum2015,TyranowskiDesbrunRAMVI} and the references therein). Deterministic geometric integrators have been successfully applied to collisionless problems in plasma physics (\cite{KrausGEMPIC,Qin2016,XiaoLiuQin2013,XiaoQinLiu2018,Xiao2015,TyranowskiDesbrunLinearLagrangians}), while stochastic variational integrators show a great promise for simulations of collisional kinetic equations (\cite{kraus2021variational,tyranowski2021stochastic}). Finally, particle simulations often require a large number of particles to achieve satisfactory accuracy (see Section~\ref{subsubsec: Maintainance of Maxwellian test particles}), which can be computationally very expensive, especially when simulations for many different values of input parameters are desired. It would be of interest to investigate the application of \emph{data-driven geometric model reduction techniques}, which could alleviate these computational costs (\cite{TyranowskiKraus2021,TyranowskiStochasticModelReduction}).

\section*{ACKNOWLEDGEMENTS} 

ZL would like to thank A. Bergmann, C. Slaby, R. Hatzky, M. Kraus, D. Coster, M. Hoelzl and P. Donnel for their inputs. Simulations have been carried out on the MPCDF and TOK cluster. 

This work has been carried out within the framework of the EUROfusion Consortium, funded by the European Union via the Euratom Research and Training Programme (Grant Agreement No 101052200 — EUROfusion). Views and opinions expressed are however those of the author(s) only and do not necessarily reflect those of the European Union or the European Commission. Neither the European Union nor the European Commission can be held responsible for them.

\newpage

\newpage
\appendix
\section{Other high-order SDE schemes for Stratonovich form}
\label{app:schemes}
\begin{enumerate}
    \item A four-stage scheme (referred to as ``CL") with $s=4$, $p=2$ (Eq. 56 of \cite{burrage1996high}) as
    \begin{eqnarray} 
         (a_{ij})&=& \scriptsize
                 \begin{bmatrix} 0 & 0 & 0 &0 \\ \frac{1}{2} & 0 & 0 & 0 \\ 
                0 & \frac{1}{2} & 0 & 0 \\ 0 & 0 & 1 & 0 \end{bmatrix}    \\
        (b_{ij,1})&=& \scriptsize
                \begin{bmatrix}
                0  &  0  & 0  & 0 \\
                -0.7242916356 & 0 & 0 & 0\\
                0.4237353406  & -0.1994437050 & 0 & 0 \\
                -1.578475506  & 0.840100343   &1.738375163 &0  &  \\
            \end{bmatrix} \\
        (b_{ij,2})&=& \scriptsize
                \begin{bmatrix}
                0 &0& 0& 0 \\
                2.702000410 & 0 & 0 & 0 \\
                1.757261649 & 0 & 0 & 0\\
                -2.918524118 & 0 & 0 & 0
            \end{bmatrix} \\
        (\alpha_i) &=& \scriptsize
                \begin{bmatrix} \frac{1}{6} & \frac{1}{3}&\frac{1}{3}&\frac{1}{6}  \end{bmatrix}  \\
        (\beta_{j,1})&=& \scriptsize
                 \begin{bmatrix} -0.7800788474 & 0.07363768240 & 1.486520013 & 0.2199211524  \end{bmatrix}   \\
        (\beta_{j,2})&=& \scriptsize
                 \begin{bmatrix} 1.693950844 & 1.636107882 & -3.024009558 & -0.3060491602 \end{bmatrix}   
    \end{eqnarray}
    \item A five-stage scheme with $s=5$, $p=2$ (``G5'') as
    \begin{eqnarray} 
        (a_{ij})=  \scriptsize\begin{bmatrix} 
                0 &0 &0 &0 &0 \\
                0.52494822322232 &0 &0 &0 &0 \\
                0.07167584568902 &0.27192330512685  & 0 &0 &0 \\
                0.13408162649312 &0.24489042208103  &-0.02150276857782 & 0 & 0 \\
                -.07483338680171 &-.07276896351874  &.55202897082453 &-.50752343840006 & 0 
        \end{bmatrix}  \\ 
        (b_{ij,1})=\scriptsize\begin{bmatrix} 
                0& 0& 0& 0& 0\\
                0.52494822322232 & 0& 0& 0& 0\\
                .49977623528582  &-.14576793502675 &0 &0 &0 \\
                .60871134749146  &.58291821365556  &-.94596532788804 & 0 &0 \\
                -.04005606091567 & -.22719654397712& -.12926284222120& .42881625288868 &0
        \end{bmatrix} \\
        (b_{ij,2}) =\scriptsize\begin{bmatrix} 
                0& 0& 0& 0& 0 \\
                0& 0& 0& 0& 0& \\
                -.23101439602069& .59278042710702& 0& 0& 0\\
                -.54946055077234& .86811263829203& .06772607159055& 0& 0\\
                .03847082280344& -.16953882944054& .88387761274601& -.85833118389518& 0
        \end{bmatrix} 
    \end{eqnarray}
    \begin{eqnarray}\scriptsize
        (\alpha_{j})^T = 
        \begin{bmatrix} 
            -5.60958180689351 \\ -0.67641638321828 \\ -5.44025143434789 \\ 8.76396506407891 \\ 3.96228456038077
        \end{bmatrix},
        (\beta_{j,1})^T= 
        \begin{bmatrix} 
            6.68050246229861 \\ 0 \\  4.28273528343281 \\ -3.25408735237225 \\ -6.70915039335930
        \end{bmatrix},
        (\beta_{j,2})^T=
        \begin{bmatrix} 
            1.90494977554482 \\ -1.90494977554482 \\ 0 \\ 0 \\ 0
        \end{bmatrix} 
    \end{eqnarray}
    
\end{enumerate}

\end{document}